\documentclass[preprint,showpacs,preprintnumbers,amsmath,amssymb]{revtex4}
\usepackage{graphicx}% Include figure files
\usepackage{dcolumn}% Align table columns on decimal point
\usepackage{bm}% bold math
\newcommand{\comment}[1]{}
\newcommand{\bea}{\begin{eqnarray}}
\newcommand{\eea}{\end{eqnarray}}
\begin{document}

\title[Classification of equation]
{Classification of Lie point symmetries for quadratic Li$\acute{\textbf{e}}$nard type equation $\ddot{x}+f(x)\dot{x}^2+g(x)=0$}

\author{Ajey K. Tiwari$^1$}
\author{S. N. Pandey$^1$}
\email{snp@mnnit.ac.in (S. N. Pandey)}
\author{M. Senthilvelan$^2$}
\email{ velan@cnld.bdu.ac.in (M. Senthilvelan)}
\author{M. Lakshmanan$^2$}
\email{ lakshman@cnld.bdu.ac.in (M. Lakshmanan)}
\affiliation{%
$^1$ Department of Physics, Motilal Nehru National Institute of
Technology, Allahabad - 211 004, India\\$^2$ Centre for Nonlinear Dynamics, School of Physics, Bharathidasan University,
 Tiruchirapalli - 620 024, India}%
\date{\today} 

\begin{abstract}
In this paper we carry out a complete classification of the Lie point symmetry groups associated with the quadratic Li$\acute{e}$nard type equation, 
$\ddot {x} + f(x){\dot {x}}^{2} + g(x)= 0$, where $f(x)$ and $g(x)$ are arbitrary functions of $x$. The symmetry analysis gets divided into two cases, $(i)$ the maximal (eight parameter) symmetry group and $(ii)$ non-maximal (three, two and one parameter) symmetry groups. We identify the most general form of the quadratic Li$\acute{e}$nard equation in each of these cases. In the case of eight parameter symmetry group, the identified general equation becomes linearizable as well as isochronic. We present specific examples of physical interest. For the nonmaximal cases, the identified equations are all integrable and include several physically interesting examples such as the Mathews-Lakshmanan oscillator, particle on a rotating parabolic well, etc. We also analyse the underlying equivalence transformations.
 \end{abstract}

\maketitle
\section{Introduction}
\label{sec1}
Ordinary differential equations (ODEs), especially nonlinear ones, are very useful in the formulation of fundamental natural laws and technological problems for a long time. The Lie algebraic properties of these equations are one of the basic aspects which got attention after Lie's initial work in which he  discovered that all the integration methods for ODEs can be obtained from his theory \cite{olver:1986,bluman:1989,Hydon:2000,cant:2002,ibra:1999,step,hill,baum,euler,lak}. Lie gave a classification of ODEs in terms of their symmetry groups, thereby identified the full set of equations which could be integrated or reduced to lower order equations by his method. Since then several contributions have been made on symmetry group classification of ODEs \cite{snp1,maho,maho1,leach,leach2,govinder,gat}. In particular it has been shown that any second order nonlinear ODE which admits eight parameter Lie point symmetries is linearizable to free particle equation through point transformations \cite{ibra:1999}.

Recently, Pandey \textit{et al.} \cite{snp1} have studied the Lie point symmetry properties of a general Li$\acute{e}$nard type equation
\begin{equation}
\ddot {x} + f(x){\dot {x}} + g(x)= 0,\,\,\,(\cdot=\frac{d}{dt})\label{intr1}
\end{equation}
where $f(x)$ and $g(x)$ are arbitrary smooth functions of $x$ and overdot denotes differentiation with respect to $t$, and identified several  interesting integrable and linearizable equations. They divided their analysis into two parts. In the first part they isolated equations that admit lesser parameter Lie point symmetries, and in the second part identified equations that admit maximal (eight parameter) Lie point symmetries. They proved the integrability of all the equations obtained in the first part either by providing the general solution or by constructing a time independent 
Hamiltonian. In the second part they discussed the linearizing transformations and solutions for all the nonlinear equations identified under this category. 

Yet another general second order nonlinear differential equation which is of high physical and mathematical interest \cite{lak,chou,saba} is of the form 
\begin{equation}
A(\ddot{x},\dot{x},x)\equiv \ddot {x} + f(x){\dot {x}}^{2} + g(x)= 0,\label{intr2}
\end{equation}
where $f(x)$ and $g(x)$ are arbitrary functions of $x$ and overdot denotes differentiation with respect to $t$. We will designate Eq. (\ref{intr2}) as a quadratic Li$\acute{e}$nard type system for convenience (corresponding to $\dot{x}^2$ term in (\ref{intr2})). For example, the one dimensional Mathews-Lakshmanan (ML) oscillator with $f(x)=-\frac{\lambda x}{1+\lambda x^2}$ and $g(x)=\frac{\omega_0^2x}{1+\lambda x^2}$ belongs to this class  \cite{mathews}. It exhibits only one Lie point symmetry even though it has been proved to be linearizable with the help of nonlocal transformations. Due to its unusual property it has been studied by many authors at the classical as well as quantum levels \cite{cari,cari1}. Another example is the motion of a particle on a rotating parabolic well with $f(x)=\frac{\lambda x}{1+\lambda x^2}$ and $g(x)=\frac{\omega_0^2x}{1+\lambda x^2}$ \cite{gold,ml1,nay}. Several theorems on the isochronous cases belonging to (\ref{intr2}) exist in the literature \cite{chou,saba,chou2,chou3}. Even though the integrability properties of the general form of Eq. (\ref{intr2}) has been discussed in the literature and it can be integrated in the form of a quadrature, its general properties have been hardly studied. In this paper, we systematically identify and classify all the equations, belonging to class (\ref{intr2}), which admit one, two, three and eight parameter symmetry groups, from a group theoretic point of view and explore certain interesting properties associated with them.

The main objective of this paper is to carry out a detailed Lie point symmetry analysis of Eq. (\ref{intr2}). In this way we study the  linearizable, and integrable (but not linearizable by point transformations) cases separately. Firstly, we consider the linearizable case and find the general form of Eq. (\ref{intr2}) for which it admits eight point symmetry generators. The general form of Eq. (\ref{intr2}) in this case turns out to be
\begin{equation}
 \label{intr3}
 \ddot{x}+f(x)\,{\dot{x}}^{2}+g_{1} {e^{-\int{f(x)dx}}} \int{e^{\int{f(x)dx}}dx}+g_{2}e^{-\int{f(x)dx}}=0,
\end{equation}
 where $g_1$ and $g_2$ are constant parameters. Apart from this we also show that the system (\ref{intr2}) additionally follows the isochronous condition, 
\begin{eqnarray}
  g_{x}+f\,g=g_1,\label{intr4}
\end{eqnarray}
where $g_1$ is a constant and subscript denotes differentiation with respect to $x$. We prove the significant result that (\ref{intr4}) implies (\ref{intr2}). We also discuss some specific examples of physical interest belonging to this class.

Secondly, we consider the integrable cases of Eq. (\ref{intr2}) with lesser parameter symmetries. The general form of the equations which show three and two parameter symmetry generators, respectively, are as follows:
\begin{eqnarray}
&&(i)\,\,  \ddot{x}+f(x)\,\dot{x}^2+g_{1}e^{-\int{fdx}}{\Bigl(\lambda_1+\int{e^{\int{fdx}}}dx\Bigl)}^{-3}=0, \label{intr5a}\\
&&(ii)\,\, \ddot{x}+f(x)\,\dot{x}^2+g_{1}e^{-\int{fdx}}{\Bigl(\lambda_1+\int{e^{\int{fdx}}}dx\Bigl)}^{1-\lambda_2}=0,\,\,\,\lambda_2\neq1,\,4,\label{intr5b}
\end{eqnarray}
where $g_1,\,\lambda_1$ and $\lambda_2$ are constants. 

The rest of the cases of Eq. (\ref{intr2}) correspond to one parameter Lie point symmetry group. In fact, ML oscillator  with $f(x)=-\frac{\lambda \,x}{1+\lambda\, x^2}, \,g(x)=\frac{\omega_0^2\,x}{1+\lambda \,x^2}$ belongs to this case. To ensure the existence of integrability of such systems, one can associate more general symmetries such as the $\lambda$-symmetries \cite{muri1,bhu}. 

The plan of the paper is as follows. In Sec. II we deduce the determining equations for the infinitesimal symmetries. The general form of the 
equation is obtained for the maximal symmetry group in Sec. III and its symmetry generators corresponding to $sl(3,R)$ algebra are deduced. The isochrnocity of the system is verified in Sec IV. Some physically interesting examples of this class of equations are discussed in Sec V. Sec. VI deals with the nonmaximal case and the associated equations are derived here. The symmetry property with $f(x)$ or $g(x)$ as zero is discussed in Sec. VII. The equivalence transformations are investigated in Sec. VIII. Finally, conclusion is given in Sec. IX.

\section{Determining equations for the infinitesimal symmetries}
\label{sec2}
We consider the quadratic Li$\acute{e}$nard type system of the form (\ref{intr2}). We will assume $f(x)\neq0$ and $g(x)\neq0$ to start with. Let Eq. (\ref{intr2}) be invariant under an one parameter group of symmetry transformations,
\begin{eqnarray}
&&T=t+\varepsilon \,\xi(t,x)+O(\epsilon^{2})\nonumber\\
&&X=x+\varepsilon \,\eta(t,x)+O(\epsilon^{2}),\quad \epsilon \ll 1.\label{a2}
\end{eqnarray}
An operator $G$ given by 
\begin{eqnarray}
G(t,x)=\xi(t,x)\,\frac{\partial }{\partial t}+\eta(t,x)\,\frac{\partial }{\partial x}\label{a3}
\end{eqnarray}
is said to be an infinitesimal generator of the one parameter Lie point symmetry group of transformations for Eq. (\ref{intr2}) iff
\begin{equation}
G^{(2)}(A)\mid_{A=0}=0,\label{a4}
\end{equation}
or equivalently
\begin{equation}
\Bigl(\xi\, \frac{\partial A}{\partial t}+\eta \,\frac{\partial A}{\partial x}+\eta^{(1)} \frac{\partial A}{\partial \dot{x}}+\eta^{(2)} \frac{\partial A}{\partial \ddot{x}}\Bigl)\,\mid_{A=0}=0,\label{a5}
\end{equation}
where
\begin{eqnarray}
G^{(2)}=G^{(1)}+\eta^{(2)}\frac{\partial}{\partial \ddot{x}},\quad G^{(1)}=G+\eta^{(1)}\frac{\partial}{\partial \dot{x}},\nonumber
\end{eqnarray}
\begin{eqnarray}
\eta^{(2)}=\frac{d\eta^{(1)}}{dt}-\ddot{x}\,\frac{d\xi}{dt},\quad \eta^{(1)}=\frac{d\eta}{dt}-\dot{x}\,\frac{d\xi}{dt}\label{a6}
\end{eqnarray}
and
\begin{equation}
\frac{d}{dt}=\frac{\partial}{\partial t}+\dot{x}\,\frac{\partial}{\partial x}.\label{a7}
\end{equation}

Substituting Eqs. (\ref{a6}) and (\ref{a7}) in Eq. (\ref{a5}) and equating different powers of $\dot{x}^m,\,m=0,\,1,\,2,\,3,$ to zero, we obtain the determining equations,
\begin{eqnarray}
&&\xi_{xx}-f\xi_{x}=0,\label{a8} \\
&&\eta_{xx}+f\eta_{x}+\eta f_{x}-2\,\xi_{tx}=0, \label{a9}\\
&&2\,\eta_{tx}+2\,f\eta_{t}-\xi_{tt}+3\,g\,\xi_{x}=0, \label{a10}\\
&&\eta_{tt}+\eta \,g_{x}-g\,\eta_{x}+2\,g\,\xi_{t}=0,\label{a11}
\end{eqnarray}
where subscripts denote partial derivatives. Solving Eq. (\ref{a8}) we get
\begin{equation}
\xi=b(t)\Im(x)+a(t), \label{a12}                    
\end{equation}
where $\Im (x) =\int{F(x)dx}$ and $F(x)=e^{\int{f(x)dx}}$. Here $a(t)$ and $b(t)$ are arbitrary functions of $t$. Substituting $\xi_{tx}=\dot{b}F$ from (\ref{a12}) into (\ref{a9}) and then integrating it twice with respect to $x$, we get
\begin{equation}
\eta =2\,\dot{b}\,G_1(x)+c(t)\,G_2(x)+d(t)\,G_3(x),\label{a16}
\end{equation}
where $G_1(x)=\frac{\int{\Im (x) F(x)dx}}{F(x)},\,\,G_2(x)=\frac{\Im(x)}{F(x)},\,\,G_3(x)=\frac{1}{F(x)}$ and $c(t)$ and $d(t)$ are arbitrary functions of $t$. Now with the above forms of $\xi$ and $\eta$, Eqs. (\ref{a10}) and (\ref{a11}) can be written as
\begin{eqnarray}
2\,f(2\,\ddot{b}\,G_{1}+\dot{c}\,G_{2}+\dot{d}\,G_{3})+4\,\ddot{b}\,G_{1x}+2\,\dot{c}\,G_{2x}+2\,\dot{d}\,G_{3x}-\ddot{b}\,\Im-\ddot{a}+3\,g\,b\,F=0,\label{a16.3}
\end{eqnarray}
and
\begin{eqnarray}
g_{x}\,(2\,\dot{b}\,G_{1}+{c}\,G_{2}+{d}\,G_{3})-g\,(2\,\dot{b}\,G_{1x}+{c}\,G_{2x}+{d}\,G_{3x}-2\,\dot{a}-2\,\dot{b}\,\Im)\nonumber\\
+2\,\dddot{b}\,G_{1}+\ddot{c}\,G_{2}+\ddot{d}\,G_{3}=0.\label{a16.4}
\end{eqnarray}

In analyzing the above system of Eqs. (\ref{a16.3}) and (\ref{a16.4}), we can distinguish two separate cases.\\
$(i)\, Case\, 1: b\neq0:$ In this case we can write Eq. (\ref{a16.3}) as 
\begin{eqnarray}
g=-\frac{2\,f(2\,\ddot{b}\,G_{1}+\dot{c}\,G_{2}+\dot{d}\,G_{3})+4\,\ddot{b}\,G_{1x}+2\,\dot{c}\,G_{2x}+2\,\dot{d}\,G_{3x}-\ddot{b}\,\Im-\ddot{a}}{3\,b\,F}.\label{a17}
\end{eqnarray}
Substituting $f$, the above form of $g$ and ${G_i}^{'}s,i=1,2,3,$ into (\ref{a16.4}) and equating the coefficients of various independent functions of $x$, a set of determining equations can be obtained. Solving the resultant determining equations one can get the symmetry functions $a(t),\,b(t),\,c(t)$ and $d(t)$ which in turn fix the form of the associated symmetries. \\
$(ii)\, Case\, 2: b=0:$ In this case Eqs. (\ref{a16.3}) and (\ref{a16.4}) are simplified to 
\begin{eqnarray}
2f(\dot{c}\,G_{2}+\dot{d}\,G_{3})+2\,(\dot{c}\,G_{2x}+\dot{d}\,G_{3x})-\ddot{a}=0,\label{a17.21}
\end{eqnarray}
and \begin{eqnarray}
g_{x}\,({c}\,G_{2}+{d}\,G_{3})-g\,({c}\,G_{2x}+{d}\,G_{3x}-2\,\dot{a})+\ddot{c}\,G_{2}+\ddot{d}\,G_{3}=0.\label{a17.22}
\end{eqnarray}
Substituting the forms of $G_2$ and $G_3$ in Eqs. (\ref{a17.21}) and (\ref{a17.22}) one can get the form of $g$ as well as the determining equations for the symmetry functions. Solving the associated determining equations, one can identify the corresponding symmetries.

Consequently, we will investigate the two cases, $(i)\, b\neq0$ and $(ii)\, b=0$, separately in sections III-V and VI, respectively,  and show that the $ b\neq0$ case corresponds to maximal (eight parameter) symmetry group, while the $ b=0$ case corresponds to nonmaximal (three, two, one parameter) symmetry group of transformations. 

\section{General form of the equation for $b\neq0$ case - Eight Parameter Symmetries}
\label{sec3}
We now consider the case $b\neq0$ in Eqs. (\ref{a16.3}) and (\ref{a16.4}). For the sake of generality we consider $f=f(x)$. Now substituting the values of ${G_{i}}^{'}s,i=1,2,3,$ (given below Eq. (\ref{a16})) and their derivatives in terms of $f(x)$ in Eq. (\ref{a17}) we arrive at 
\begin{equation}
g=g_{1} {e^{-\int{f(x)dx}}} \int{e^{\int{f(x)dx}}dx}+g_{2}e^{-\int{f(x)dx}},\label{b2}
\end{equation}
where $g_{1}=-\frac{\ddot{b}}{b}=\text{constant}$ and $g_{2}=-\frac{2\,\dot{c}-\ddot{a}}{3\,b}=\text{constant}$ as $g$ is a function of $x$ only and is free from $t$.\\
Now for the above form of $g(x)$ Eq. (\ref{intr2}) can be written as
\begin{equation}
 \ddot{x}+f(x)\,{\dot{x}}^{2}+g_{1} {e^{-\int{f(x)dx}}} \int{e^{\int{f(x)dx}}dx}+g_{2}e^{-\int{f(x)dx}}=0.\label{b3}
\end{equation}
To explore the symmetry group of the above Eq. (\ref{b3}), we substitute the values of ${G_{i}}^{'}s$ and their derivatives as well as $g$ in (\ref{a16.4}), and equate the resultant coefficients of various independent functions of $x$ to zero. Consequently we get the following set of determining equations for the functions $a(t),b(t),c(t)$ and $d(t)$, that is
\begin{eqnarray}
\ddot{b}+g_{1}\,b&=&0,\quad \ddot{a}-2\,\dot{c}-3\,g_{2}\,b=0,\nonumber\\
\ddot{c}+2\,g_{1}\,\dot{a}&=&0,\quad \ddot{d}+g_{1}\,d+2\,g_{2}\,\dot{a}-g_{2}\,c=0.\label{b4}
\end{eqnarray}
Note that $g_1$ and $g_2$ are system parameters fixed by the form of $g(x)$ in (\ref{b2}). Solving the above system of equations consistently, we can express the form of the functions $a(t),b(t),c(t)$ and $d(t)$ as
\begin{eqnarray}
 a\left(t\right)&=&a_1-\frac{1}{2\,g_1}\left(c_2\sin \left( 2\, \sqrt{g_{{1}}}t \right) +c_3\cos \left( 2\, \sqrt{g_{{1}}}t \right)-2\,g_2\,b_1\sin \left(  \sqrt{g_{{1}}}t \right)-2\,g_2\,b_2\cos \left(  \sqrt{g_{{1}}}t \right)\right),\nonumber\\
b \left( t \right) &=&b_{{1}}\sin \left(  \sqrt{g_{{1}}}t \right) +b_{{2}}\cos \left(  \sqrt{g_{{1}}}t \right),\nonumber\\
c \left( t \right)&=&c_1-\frac{1}{2\,\sqrt{g_1}}\left(c_2\cos \left( 2\, \sqrt{g_{{1}}}t \right) -c_3\sin \left( 2\, \sqrt{g_{{1}}}t \right)-4\,g_2\,b_1\cos \left(  \sqrt{g_{{1}}}t \right)+4\,g_2\,b_2\sin \left(  \sqrt{g_{{1}}}t \right)\right),\nonumber\\
d \left( t \right) &=&d_1\sin \left(  \sqrt{g_{{1}}}t \right)+d_2\cos \left(  \sqrt{g_{{1}}}t \right)+\frac{g_2}{2\,g_1^{3/2}}\left(2\,c_1\sqrt{g_1}-c_2\cos \left( 2\, \sqrt{g_{{1}}}t \right)+c_3\sin \left( 2\, \sqrt{g_{{1}}}t \right)\right),\nonumber\\
\label{b5}
\end{eqnarray}
where $a_1,\,b_1,\,b_2,\,c_1,\,c_2,\,c_3,\,d_1$ and $d_2$ are arbitrary constants, which indeed constitute the eight symmetry parameters corresponding to the maximal symmetry group. Thus we can conclude that the specific differential Eq. (\ref{b3}) admits the maximal symmetry group. The corresponding infinitesimal symmetries (which follow from (\ref{a12}) and (\ref{a16})) are
\begin{eqnarray}
\xi&=&a_{1}+b_1\,\frac{\sin \left( \sqrt{g_{{1}}}t \right)(g_1 \int \!{e}^{\int \!f \left( x \right) {dx}}{dx}+g_2)}{g_1}+b_2\,\frac{\cos \left( \sqrt{g_{{1}}}t \right)(g_1 \int \!{e}^{\int \!f \left( x \right) {dx}}{dx}+g_2)}{g_1}\nonumber\\
&&-c_2\,\frac{\sin \left( \sqrt{2\,g_{{1}}}t \right)}{2\,g_1}-c_3\,\frac{\cos \left( \sqrt{2\,g_{{1}}}t \right)}{2\,g_1},\nonumber\\
\eta&=&b_1\,\frac{{e}^{-\int \!f \left( x \right) {dx}}\left( \int \!{e}^{\int \!f \left( x \right) {dx}}{dx} \right)\cos \left(  \sqrt{g_{{1}}}t \right)(g_1 \int \!{e}^{\int \!f \left( x \right) {dx}}{dx}+2\,g_2)}{\sqrt{g_1}}\nonumber\\
&&-b_2\,\frac{{e}^{-\int \!f \left( x \right) {dx}}\left( \int \!{e}^{\int \!f \left( x \right) {dx}}{dx} \right)\sin \left(  \sqrt{g_{{1}}}t \right)(g_1 \int \!{e}^{\int \!f \left( x \right) {dx}}{dx}+2\,g_2)}{\sqrt{g_1}}\nonumber\\
&&+c_1\,\frac{{e}^{-\int \!f \left( x \right) {dx}}(g_1 \int \!{e}^{\int \!f \left( x \right) {dx}}{dx}+g_2)}{g_1}-c_2\,\frac{{e}^{-\int \!f \left( x \right) {dx}}\cos \left( 2\, \sqrt{g_{{1}}}t \right)(g_1 \int \!{e}^{\int \!f \left( x \right) {dx}}{dx}+g_2)}{2g_1^{3/2}}\nonumber\\
&&+c_3\,\frac{{e}^{-\int \!f \left( x \right) {dx}}\sin \left( 2\, \sqrt{g_{{1}}}t \right)(g_1 \int \!{e}^{\int \!f \left( x \right) {dx}}{dx}+g_2)}{2g_1^{3/2}}+d_{{1}}\sin \left(  \sqrt{g_{{1}}}t \right){e}^{-\int \!f \left( x \right) {dx}} \nonumber\\
&&+d_{{2}}\cos \left(  \sqrt{g_{{1}}}t \right){e}^{-\int \!f \left( x \right) {dx}}. \label{b6}
\end{eqnarray}
The associated infinitesimal generators read
\begin{eqnarray}
X_{1}&=&\frac{\partial }{\partial t},\nonumber\\
X_{2}&=&\frac{\sin \left( \sqrt{g_{{1}}}t \right)(g_1 \int \!{e}^{\int \!f \left( x \right) {dx}}{dx}+g_2)}{g_1}\,\frac{\partial}{\partial t}\nonumber\\
&&+\frac{{e}^{-\int \!f \left( x \right) {dx}}\left( \int \!{e}^{\int \!f \left( x \right) {dx}}{dx} \right)\cos \left(  \sqrt{g_{{1}}}t \right)(g_1 \int \!{e}^{\int \!f \left( x \right) {dx}}{dx}+2g_2)}{\sqrt{g_1}}\,\frac{\partial}{\partial x},\nonumber\\
X_{3}&=&\frac{\cos \left( \sqrt{g_{{1}}}t \right)(g_1 \int \!{e}^{\int \!f \left( x \right) {dx}}{dx}+g_2)}{g_1}\,\frac{\partial}{\partial t}\nonumber\\
&&-\frac{{e}^{-\int \!f \left( x \right) {dx}}\left( \int \!{e}^{\int \!f \left( x \right) {dx}}{dx} \right)\sin \left(  \sqrt{g_{{1}}}t \right)(g_1 \int \!{e}^{\int \!f \left( x \right) {dx}}{dx}+2g_2)}{\sqrt{g_1}}\,\frac{\partial}{\partial x},\nonumber\\
X_{4}&=&\frac{{e}^{-\int \!f \left( x \right) {dx}}(g_1 \int \!{e}^{\int \!f \left( x \right) {dx}}{dx}+g_2)}{g_1}\,\frac{\partial}{\partial x},\nonumber\\
X_{5}&=&-\frac{\sin \left( \sqrt{2\,g_{{1}}}t \right)}{2\,g_1}\,\frac{\partial}{\partial t}-\frac{{e}^{-\int \!f \left( x \right) {dx}}\cos \left( 2\, \sqrt{g_{{1}}}t \right)(g_1 \int \!{e}^{\int \!f \left( x \right) {dx}}{dx}+g_2)}{2g_1^{3/2}}\,\frac{\partial}{\partial x},\nonumber\\
X_{6}&=&-\frac{\cos \left( \sqrt{2\,g_{{1}}}t \right)}{2\,g_1}\,\frac{\partial}{\partial t}+\frac{{e}^{-\int \!f \left( x \right) {dx}}\sin \left( 2\, \sqrt{g_{{1}}}t \right)(g_1 \int \!{e}^{\int \!f \left( x \right) {dx}}{dx}+g_2)}{2g_1^{3/2}}\,\frac{\partial}{\partial x},\nonumber\\
X_{7}&=&\sin \left(  \sqrt{g_{{1}}}t \right){e}^{-\int \!f \left( x \right) {dx}}\,\frac{\partial}{\partial x},\nonumber\\
X_{8}&=&\sin \left(  \sqrt{g_{{1}}}t \right){e}^{-\int \!f \left( x \right) {dx}}\,\frac{\partial}{\partial x}.\label{b7}
\end{eqnarray}
One can easily check that these eight generators lead to the $sl(3,R)$ algebra.

\section{Isochronous condition, Linearizability and the nature of solution of Eq. (\ref{b3})}
\label{sec4}
\subsection{Isochronicity condition}
A dynamical system is called isochronous if it features in its phase space an open, fully-dimensional region where all its solutions are periodic in all its degrees of freedom with the same, fixed, period \cite{calo}. The linear harmonic oscillator is the prototype of an isochronous system and all other isochronous systems are isoperiodic with the harmonic oscillator. Eq. (\ref{intr2}) can be mapped on to the linear harmonic oscillator, 
\begin{equation}
 \ddot{X}+\omega_0^2\,X=0,\label{c1}   
\end{equation}
with the following invertible transformation
\begin{equation}
 X=h(x),\label{c2}
\end{equation}
where $h(x)$ is a function of $x$ alone, provided for suitable choice of $f(x)$ and $g(x)$ Eq. (\ref{intr2}) is isochronic. Now substituting the derivatives of $X$ in equation (\ref{c1}) we get
\begin{equation}
 \ddot{x}+\frac{ h^{''}(x)}{ h^{'}(x)}\,\dot{x}^2+\omega_0^2\, \frac{ h(x)}{ h^{'}(x)}=0,\label{c3} 
\end{equation}
where prime denotes differentiation with respect to $x$. Comparing Eq. (\ref{c3}) with (\ref{intr2}) we get the following two relations
\begin{eqnarray}
\frac{ h^{''}(x)}{ h^{'}(x)}=f(x),\label{c4}\\
\omega_0^2\, \frac{ h(x)}{ h^{'}(x)}=g(x).\label{c5}
\end{eqnarray}
From Eq. (\ref{c4}) one can get
\begin{equation}
h(x)=h_1\,\int{e^{\int {f(x)dx}}dx}+h_2,\label{c6}
\end{equation} 
where $h_1$ and $h_2$ are integration constants. Now substituting (\ref{c6}) and its first derivative into Eq. (\ref{c5}) we can obtain
\begin{equation}
 h(x)=\frac{h_1}{\omega_0^2}\,g(x)e^{\int{f(x)dx}}.\label{c7}
\end{equation}
Comparing the forms of $h(x)$ from Eq. (\ref{c6}) and (\ref{c7}) we obtain the condition
\begin{equation}
 h_1\,g(x)\,e^{\int{f(x)dx}}=\omega_0^2\,\Bigl(h_1\int{e^{\int{f(x)dx}}dx}+h_2\Bigl).\label{c8}
\end{equation}
Differentiating both sides of (\ref{c8}) with respect to $x$ and integrating the resultant equation we find
\begin{equation}
g=g_{1} {e^{-\int{f(x)dx}}} \int{e^{\int{f(x)dx}}dx}+g_{2}e^{-\int{f(x)dx}},\label{c10}
\end{equation}
where $\omega_0^2$ has been replaced by $g_1$ for analogy. Eq. (\ref{c10}) is exactly the same as the one we obtained with the help of symmetry method, see Eq. (\ref{b2}). Now inverting the relation $X=h(x)$ in Eq. (\ref{c2}), we find that for the above form of $g$, the solution is isochronous. Thus we can conclude that the periodic solutions of Eq. (\ref{b3}) are all isochronous.

\subsection{Linearizability Condition}
It has been proved that the linearization of a scalar second order ODE, $\ddot{x}+f(t,x,\dot{x})=0$, via point transformation has the cubic in first 
derivative \cite{ibra:1999}, that is
\begin{equation}
\ddot{x}=P(t,x)\dot{x}^{3}+Q(t,x)\dot{x}^{2}+R(t,x)\dot{x}+S(t,x),\label{c11}
\end{equation}
with the coefficients $A,\,B,\,C$ and $D$ satisfying the following two invariant conditions,
\begin{eqnarray}
3P_{tt}+3P_{t}R-3P_{x}S+3PR_{t}+R_{xx}-6PS_{x}+QR_{x}-2QQ_{t}-2Q_{tx}=0,\nonumber\\
6P_{t}S-3Q_{x}S+3PS_{t}+Q_{tt}-2R_{tx}-3QS_{x}+3S_{xx}+2RR_{x}-RQ_{t}=0,\label{c12}
\end{eqnarray}
where the suffix refers to partial derivatives. On comparing (\ref{intr2}) with (\ref{c11}), we find the condition of linearizability from equation (\ref{c12}) as
\begin{equation}
 g_{xx}+fg_{x}+gf_{x}=0.\label{c13}
\end{equation}
The integration of the above equation gives the form of $g(x)$ as
\begin{equation}
g=g_{1} {e^{-\int{f(x)dx}}} \int{e^{\int{f(x)dx}}dx}+g_{2}e^{-\int{f(x)dx}}.\label{c14}
\end{equation}
So we see that the linearizability criterion also suggests the same form of $g$ as the one we obtained from symmetry method and by isochronicity condition. This is in conformity with the fact that the system (\ref{b3}) admits eight Lie point symmetries as shown above.

\section{Special cases of maximal symmetry group}
\label{sec5}
In this section we consider some physically interesting special cases by fixing the form of $f(x)$ in Eq. (\ref{b3}). The determining equations will be the same as given by Eq. (\ref{b4}) and hence the forms of the functions $a(t), b(t),c(t)$ and $d(t)$ will also be the same. Then their symmetries can be obtained directly from Eq. (\ref{b6}) for appropriate choice of $f(x)$ and parameters, which leads to the infinitesimal generators. 

\subsection{$f(x)=\lambda =\text{constant}$} 
\noindent Fixing the function $f(x)$ as a constant $(=\lambda) $, we get the form of Eq. (\ref{intr2}) from Eq. (\ref{b3}) as a generalized Morse oscillator,
\begin{eqnarray}
 \ddot{x}+\lambda \, \dot{x}^{2}+\frac{\omega_0^2}{\lambda}\,(1-{e}^{-\lambda x})=0.\label{d1.1}
\end{eqnarray}
Note that in the limit $\lambda \rightarrow 0$, Eq. (\ref{d1.1}) reduces to the linear harmonic oscillator
\begin{equation}
\ddot{x}+\omega_o^2\,x=0.\label{d1.2}
\end{equation}
The form of $\xi$ and $\eta$ associated with Eq. (\ref{d1.1}) will be obtained directly from equation (\ref{b6}) and can be written as
\begin{eqnarray}
\xi \left( t,x \right) &=&a_{{1}}+b_1\,\frac{\sin\left(\omega_0 t \right)\,(e^{\lambda x}-1)}{\lambda}+b_2\,\frac{\cos\left(\omega_0 t \right)\,(e^{\lambda x}-1)}{\lambda}-c_2\,\frac{\sin\left(2\,\omega_0 t \right)}{2\,\omega_0^2}-c_3\,\frac{\cos\left(2\,\omega_0 t \right)}{2\,\omega_0^2},\nonumber\\
\eta \left( t,x \right)&=&b_1\,\frac{\omega_0\cos\left(\omega_0 t \right)\,(e^{\lambda x}-2)}{\lambda^2}-b_2\,\frac{\omega_0\sin\left(\omega_0 t \right)\,(e^{\lambda x}-2)}{\lambda^2}+c_1\,\frac{1-e^{-\lambda x}}{\lambda}\nonumber\\
&&-c_2\,\frac{\cos\left(2\,\omega_0 t \right)\,(1-e^{-\lambda x})}{2\,\lambda\,\omega_0}+c_3\,\frac{\sin\left(2\,\omega_0 t \right)\,(1-e^{-\lambda x})}{2\,\lambda\,\omega_0}\nonumber\\
&&+d_1\,\sin\left(\omega_0 t \right)e^{-\lambda x}+d_2\,\cos\left(\omega_0 t \right)e^{-\lambda x}.\label{d1.3}
\end{eqnarray}
The corresponding symmetry generators thus turn out to be
\begin{eqnarray}
X_{1}&=&\frac{\partial}{\partial t},\,\,\,\,\,X_{2}=\frac{\sin\left(\omega_0 t \right)\,(e^{\lambda x}-1)}{\lambda}\,\frac{\partial}{\partial t}+\frac{\omega_0\cos\left(\omega_0 t \right)\,(e^{\lambda x}-2)}{\lambda^2}\,\frac{\partial}{\partial x},\nonumber\\
X_{3}&=&\frac{\cos\left(\omega_0 t \right)\,(e^{\lambda x}-1)}{\lambda}\,\frac{\partial}{\partial t}-\frac{\omega_0\sin\left(\omega_0 t \right)\,(e^{\lambda x}-2)}{\lambda^2}\,\frac{\partial}{\partial x},\,\,\,X_{4}=\frac{1-e^{-\lambda x}}{\lambda}\,\frac{\partial}{\partial x},\nonumber\\
X_{5}&=&-\frac{\sin\left(2\,\omega_0 t \right)}{2\,\omega_0^2}\,\frac{\partial}{\partial t}-\frac{\cos\left(2\,\omega_0 t \right)\,(1-e^{-\lambda x})}{2\,\lambda\,\omega_0}\,\frac{\partial}{\partial x},\nonumber\\
X_{6}&=&-\frac{\cos\left(2\,\omega_0 t \right)}{2\,\omega_0^2}\,\frac{\partial}{\partial t}+\frac{\sin\left(2\,\omega_0 t \right)\,(1-e^{-\lambda x})}{2\,\lambda\,\omega_0}\,\frac{\partial}{\partial x},\nonumber\\
X_{7}&=&\sin\left(\omega_0 t \right)e^{-\lambda x}\,\frac{\partial}{\partial x},\,\,\,\,X_{8}=\cos\left(\omega_0 t \right)e^{-\lambda x}\,\frac{\partial}{\partial x}.\label{d1.4}
\end{eqnarray}

One can write down the solution of equation (\ref{d1.1}) by transforming it to the linear harmonic oscillator equation as mentioned above and it is  found to be
\begin{eqnarray}
x=\frac{1}{\lambda}\ln{(1-\lambda A\sin({\omega_0t+\delta)})},\,\,\,\,\,0\leq A<\frac{1}{\lambda}.\label{d1.5}
\end{eqnarray}
Here $A$ and $\delta$ are arbitrary constants. For $A< 1$ we have physically acceptable (isochronous) periodic solutions with frequency exactly the same as that of the linear harmonic oscillator case, $\lambda =0.$ The Hamiltonian of equation (\ref{d1.1}) can be then written as
\begin{equation}
H=\frac{1}{2\lambda^2}\, e^{-2\lambda x}p^2+\frac{1}{2}\,\omega_0^2\,(1-e^{\lambda x})^2,\label{d1.6}
\end{equation}
where $p=\lambda ^2e^{2\lambda x}\dot{x}$ is the conjugate momentum.

The phase portrait underlying Eq. (\ref{d1.1}) corresponding to the Hamiltonian (\ref{d1.6}) for four different sets of $A$ values with $\lambda=0.8,\, \omega=2.6$ and $\delta=0$ is shown with four different colors in Fig.1. Note that regular motion is restricted to $-\infty <x\leq \frac{1}{\lambda}\ln{2}$. Outside this region, the solution becomes singular periodically. 
\begin{figure}[!ht]
\begin{center}
\includegraphics[width=.25\linewidth]{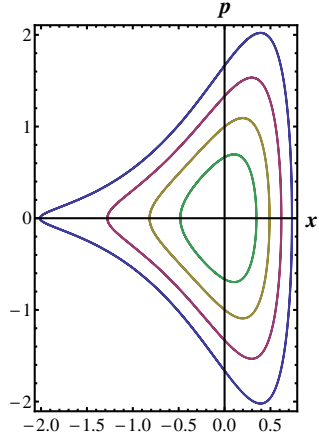}
\caption{(Color online) Phase portrait of equation (\ref{d1.1}) for different values of $A$ corresponding to the Hamiltonian (\ref{d1.6})}
\end{center}
\end{figure}

\subsection{$f(x)=-\frac{2\,\lambda}{1+\lambda \,x}$}
If we consider the form of the function $f(x)$ as $f=-\frac{2\,\lambda}{1+\lambda \,x}$, Eq. (\ref{b3}) becomes
\begin{eqnarray}
 \ddot{x}-\frac{2\,\lambda}{1+\lambda \,x}\,\dot{x}^{2}+\omega_0^2\,(x+\lambda \,{x}^{2})=0.\label{d2.1}
\end{eqnarray}
Note that in the limit $\lambda \rightarrow 0$, Eq. (\ref{d2.1}) reduces to that of a linear harmonic oscillator. The corresponding infinitesimal symmetries associated with (\ref{d2.1}) can be directly written with the help of Eq. (\ref{b6}) as
\begin{eqnarray}
\xi \left( t,x \right)&=&a_{{1}}+b_1\,\frac{x \sin \left( \omega_0 t \right)}{1+\lambda \,x}+b_2\,\frac{x \cos \left( \omega_0 t \right)}{1+\lambda \,x}-c_2\,\frac{\sin \left(2\,\omega_0 t \right)}{2\,\omega_0^2}-c_3\,\frac{\cos \left(2\,\omega_0 t \right)}{2\,\omega_0^2},\nonumber\\
\eta \left( t,x \right)&=&b_1\,\frac{\omega_0\cos \left(\omega_0 t\right)(1+2\,\lambda \,x)}{\lambda^2}-b_2\,\frac{\omega_0\sin \left(\omega_0 t\right)(1+2\,\lambda \,x)}{\lambda^2}+c_1\,(x+\lambda \,x^2)\nonumber\\
&&-c_2\,\frac{(x+\lambda \,x^2)\cos \left(2\,\omega_0t\right)}{2\,\omega_0}+c_3\,\frac{(x+\lambda \,x^2)\sin \left(2\,\omega_0t\right)}{2\,\omega_0}\nonumber\\
&&+d_1\,\sin \left(\omega_0 t \right)(1+\lambda \,x)^2+d_2\,\cos \left(\omega_0 t \right)(1+\lambda \,x)^2.
\end{eqnarray}
The corresponding symmetry generators are given by
\begin{eqnarray}
X_{1}&=&\frac{\partial}{\partial t},\,\,\,\,X_{2}=\frac{x \sin \left( \omega_0 t \right)}{1+\lambda \,x}\,\frac{\partial}{\partial t}+\frac{\omega_0\cos \left(\omega_0 t\right)(1+2\,\lambda \,x)}{\lambda^2}\,\frac{\partial}{\partial x},\nonumber\\
X_{3}&=&\frac{x \cos \left( \omega_0 t \right)}{1+\lambda \,x}\,\frac{\partial}{\partial t}-\frac{\omega_0\sin \left(\omega_0 t\right)(1+2\,\lambda \,x)}{\lambda^2}\,\frac{\partial}{\partial x},\,\,\,X_{4}=(x+\lambda \,x^2)\,\frac{\partial}{\partial x},\nonumber\\
X_{5}&=&-\frac{\sin \left(2\,\omega_0 t \right)}{2\,\omega_0^2}\,\frac{\partial}{\partial t}-\frac{(x+\lambda \,x^2)\cos \left(2\,\omega_0t\right)}{2\,\omega_0}\,\frac{\partial}{\partial x},\nonumber
\end{eqnarray}
\begin{eqnarray}
X_{6}&=&-\frac{\cos \left(2\,\omega_0 t \right)}{2\,\omega_0^2}\,\frac{\partial}{\partial t}+\frac{(x+\lambda \,x^2)\sin \left(2\,\omega_0t\right)}{2\,\omega_0}\,\frac{\partial}{\partial x},\nonumber\\
X_{7}&=&\sin \left(\omega_0 t \right)(1+\lambda \,x)^2\,\frac{\partial}{\partial x},\,\,\,\,X_{8}=\cos \left(\omega_0 t \right)(1+\lambda \,x)^2\,\frac{\partial}{\partial x}.\label{d2.3}
\end{eqnarray}
Equation (\ref{d2.1}) can be transformed to the linear harmonic oscillator equation as discussed in Sec. IV and the isochronous solution can be then written as
\begin{eqnarray}
x=\frac{A\sin{(\omega_0t+\delta)}}{1-\lambda A \sin({\omega_0t+\delta})},\,\,\,\,\, 0\leq A <\frac{1}{\lambda},\label{d2.4}
\end{eqnarray}
where $\omega_0$ and $\delta$ are constants. For $A< \frac{1}{\lambda}$, we have physically acceptable periodic solutions. The frequency of oscillation is exactly the same as that of the linear harmonic oscillator case, $\lambda=0$. The Hamiltonian of (\ref{d2.1}) can be written as
\begin{equation}
H=\frac{1}{2}(1+\lambda \,x)^4 p^2+\frac{1}{2}\,\frac{\omega_0^2\,x^2}{(1+\lambda \,x)^2},\label{d2.5}
\end{equation}
where the conjugate momentum is $p=\frac{\dot{x}}{(1+\lambda \,x)^4}$. The phase portrait of Eq. (\ref{d2.1}) for four different sets of $A$ values with $\lambda=0.8,\, \omega=1.5$ and $\delta=0$ is shown with four different colors in Fig.2. Note that $x$ is restricted to the region $-\frac{1}{\lambda} <x<\infty$ for periodic solutions without singularity. 
\begin{figure}[!ht]
\begin{center}
\includegraphics[width=.3\linewidth]{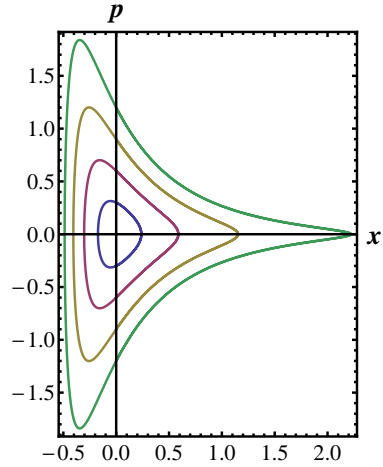}
\caption{(Color online) Phase portrait of equation (\ref{d2.1}) for different values of ${A}$ corresponding to the Hamiltonian (\ref{d2.5})}
\end{center}
\end{figure}

\section{Non-maximal symmetry: Case $b=0$}
\label{sec6}
To explore the forms of the equations having lesser parameter Lie point symmetries we consider the case $b=0$ vide Eqs. (\ref{a16.3}) and (\ref{a16.4}). Eq. (\ref{a17.21}) provides condition between the symmetry functions $a(t),\,c(t)$ and $d(t)$, whereas Eq. (\ref{a17.22}) determines the form of $g$ corresponding to the given form of $f$. 

Let us now define a function $M=2\,(\dot{c}\,G_{2}+\dot{d}\,G_{3})$. Then Eq. (\ref{a17.21}) can be rewritten as
\begin{equation}
    fM+M_{x}-\ddot{a}=0.\label{d3}
   \end{equation}
Integrating Eq. (\ref{d3}) we get
\begin{eqnarray}
MF=\ddot{a}\,{\Im (x)}+{d_1},\label{d6}
\end{eqnarray}
where $d_1$ is an integration constant. Consequently, we have
\begin{equation}
2\,\dot{c}\,G_{2}+2\,\dot{d}\,G_{3}=\ddot{a}\,\frac{\Im}{F}+\frac{d_1}{F},\label{d7}
\end{equation}
that is
\begin{equation}
 (2\,\dot{c}-\ddot{a})\,G_{2}+(2\,\dot{d}-d_1)\,G_{3}=0.\label{d8}
\end{equation}
As $G_{2}\neq0$ and $G_{3}\neq0$, which are two distinct functions of $f(x)$, we are left with two constraints
\begin{equation}
 2\,\dot{c}=\ddot{a},\quad 2\,\dot{d}=d_1.\label{d9}
\end{equation}

Next we note that Eq. (\ref{a17.22}) is a first order differential equation in $g$. For a given form of $f$ in (\ref{a17.22}) one can get $g$ which will then decide the form of (\ref{intr2}). Rewriting Eq. (\ref{a17.22}) with $\ddot{d}=0$ (from (\ref{d9})), we have
\begin{eqnarray}
g_{x}-g\,\Bigl(\frac{{c}\,G_{2x}+{d}\,G_{3x}-2\,\dot{a}}{{c}\,G_{2}+{d}\,G_{3}}\Bigl)+\frac{\ddot{c}\,G_{2}}{{c}\,G_{2}+{d}\,G_{3}}=0.\label{d10}
\end{eqnarray}
Since $g$ should be a function of $x$ alone (vide Eq. (\ref{intr2})), we choose
\begin{eqnarray}
 \frac{d}{c}=\lambda_1,\,\,\,\,\frac{2\dot{a}}{c}=\lambda_2,\,\,\,\,-\,\frac{\ddot{c}}{c}=\lambda_3,\label{d11}
\end{eqnarray}
where $\lambda_1,\lambda_2$ and $\lambda_3$ are constants. Then Eq. (\ref{d10}) can be rewritten as
\begin{eqnarray}
 g_{x}-g\,\Bigl(\frac{G_{2x}+\lambda_1G_{3x}-\lambda_2}{G_{2}+\lambda_1G_{3}}\Bigl)+\frac{\lambda_3G_{2}}{G_{2}+\lambda_1G_{3}}=0.
\label{d12}
\end{eqnarray}
Analyzing the relations $2\,\dot{c}=\ddot{a}$ (from (\ref{d9})) and $\lambda_2=\frac{2\dot{a}}{c}$ (from(\ref{d11})), we get
\begin{eqnarray}
 (\lambda_2-4)\dot{c}=0,\label{d13}
\end{eqnarray}
provided $(G_2+\lambda_1 G_3)\neq0$ or $c\,G_2+d\,G_3\neq0$. Hence we are left with three different possibilities as $(i)\,\lambda_2=4,\,\dot{c}\neq0$ and $(c\,G_2+d\,G_3)\neq0,\,(ii)\,\lambda_2\neq1,\,4,\,\dot{c}=0$ and $(c\,G_2+d\,G_3)\neq0$ and $(iii)\,(c\,G_2+d\,G_3)=0$. In the following subsections we consider all the three cases one by one and will show that the above cases $(i),\,(ii)$ and $(iii)$ correspond to three, two and one parameter Lie point symmetries,  respectively.   

\subsection{Three parameter symmetry (Case \,(i)\,$\lambda_2=4,\,\dot{c}\neq0$ \text{and} $(c\,G_2+d\,G_3)\neq0$)}
\label{sec6.1}
Solving Eqs. (\ref{d9}) and (\ref{d11}) for $\lambda_2=4$ with $\dot{c}\neq0$ we get $\lambda_3=0$ and the explicit forms for the functions $a,\,c$ and $d$ turn out to be
\begin{eqnarray}
a=a_1+\frac{2}{\lambda_1}\,\bigl(\frac{d_1}{4}\,t^2+d_2t\bigl),\,\,\,\,c=\frac{1}{\lambda_1}\bigl(\frac{d_1}{2}\,t+d_2\bigl),\,\,\,\,d=\bigl(\frac{d_1}{2}\,t+d_2\bigl),\label{d14}
\end{eqnarray}
where $a_1,\,d_1$ and $d_2$ are three arbitrary (symmetry) parameters, which lead to a three parameter Lie point symmetry group. Substituting $\lambda_2=4,\,\lambda_3=0$ and the actual forms of $G_2$ and $G_3$ in Eq. (\ref{d12}) and integrating, we obtain
\begin{equation}
g(x)=g_{1}e^{-\int{fdx}}\Bigl(\lambda_1+\int{e^{\int{fdx}}dx}\Bigl)^{-3},\label{d15}
\end{equation}
where $g_1$ is an integration constant. For this form of $g$ Eq. (\ref{intr2}) reads
\begin{eqnarray}
 \ddot{x}+f(x)\,\dot{x}^2+g_{1}e^{-\int{fdx}}\Bigl(\lambda_1+\int{e^{\int{fdx}}dx}\Bigl)^{-3}=0.\label{d16}
\end{eqnarray}
The infinitesimal symmetries associated with Eq. (\ref{d16}) are then 
\begin{eqnarray}
\xi=a_1+\frac{2}{\lambda_1}\Bigl(\frac{d_1}{4}\,t^2+d_2t\Bigl),\,\,\eta=\frac{1}{\lambda_1}\,e^{-\int{fdx}}\Bigl(\lambda_1+\int{e^{\int{fdx}}dx}\Bigl)\bigl(\frac{d_1}{2}\,t+d_2\bigl).\label{d17}
\end{eqnarray}
The corresponding infinitesimal generators can be written as
\begin{eqnarray}
 X_1&=&\frac{\partial}{\partial t},\,\,\,X_2=\frac{t^2}{2\,\lambda_1}\,\frac{\partial}{\partial t}+\frac{te^{-\int{fdx}}\Bigl(\lambda_1+\int{e^{\int{fdx}}dx}\Bigl)}{2\,\lambda_1}\frac{\partial}{\partial x},\nonumber\\
X_3&=&\frac{2\,t}{\lambda_1}\,\frac{\partial}{\partial t}+\frac{e^{-\int{fdx}}\Bigl(\lambda_1+\int{e^{\int{fdx}}dx}\Bigl)}{\lambda_1}\frac{\partial}{\partial x}.\label{d18}
\end{eqnarray}
This represents a three parameter symmetry case. The corresponding Lie algebra is 
\begin{eqnarray}
[X_1,X_2]=\frac{1}{2}X_3,\,\,[X_1,X_3]=\frac{2}{\lambda_1}X_1,\,\,[X_2,X_3]=-\frac{2}{\lambda_1}X_2.\label{d19}
\end{eqnarray}

\subsubsection*{Example:}
As a specific example of three parameter Lie point symmetries we consider the nonlinear ODE, with $f(x)=-\frac{2}{x}$ and $g_1=\lambda$ in Eq. (\ref{d16}), 
\begin{equation}
 \ddot{x}-\frac{2}{x}\,\dot{x}^2+\frac{\lambda \,x^5}{(\lambda_1\, x-1)^3}=0.\label{dEx1.1}
\end{equation}
The infinitesimal symmetries for this equation can be written from Eq. (\ref{d17}) as
\begin{equation}
 \xi=a_1+\frac{2}{\lambda_1}\bigl(\frac{d_1}{4}\,t^2+d_2\,t\bigl),\,\,\,\eta=\Bigl(\frac{\lambda_1\,x^2-x}{\lambda_1}\Bigl)\bigl(\frac{d_1}{2}\,t+d_2\bigl).\label{dEx2}
\end{equation}
The associated infinitesimal generators read
\begin{eqnarray}
X_1=\frac{\partial}{\partial t},\,\,X_2=\frac{t^2}{2\lambda_1}\,\frac{\partial}{\partial t}+\Bigl(\frac{\lambda_1x^2-x}{2\,\lambda_1}\Bigl)\,{t}\,\frac{\partial}{\partial x},\,\,X_3=\frac{2\,t}{\lambda_1}\,\frac{\partial}{\partial t}+\Bigl(\frac{\lambda_1x^2-x}{\lambda_1}\Bigl)\frac{\partial}{\partial x}.\label{dEx3}
\end{eqnarray}
The general solution of Eq. (\ref{dEx1.1}) is given by \cite{mathews} 
\begin{equation}
x(t)=\frac{\lambda_1(m\lambda_1^2 \mp \sqrt{m\lambda_1^4(m-I_1\lambda_1^2)+m^3(t-t_0)^2})}{I_1\lambda_1^6-m^2\,(t-t_0^2)},\label{dEx5}
\end{equation}
where $I_1,\,t_0$ are constants of integration and $m=I_1\lambda_1^2-\lambda$.

\subsection{Two parameter symmetry (Case\, (ii)\,$\lambda_2\neq1,\,4,\,\dot{c}=0$ and $(c\,G_2+d\,G_3)\neq0$)}
\label{sec6.2}
Again solving Eqs. (\ref{d9}) and (\ref{d11}) consistently for the choice $\lambda_2\neq4,\,\dot{c}=0$ with $(c\,G_2+d\,G_3)\neq0$, we get $\lambda_3=0$ and the explicit forms for the functions $a,\,c$ and $d$ as
\begin{eqnarray}
a=a_1+a_2t,\,\,c=\frac{2a_2}{\lambda_2},\,\,d=\frac{2a_2\lambda_1}{\lambda_2},\label{d20}
\end{eqnarray}
where $a_1$ and $a_2$ two arbitrary constants which corresponds to two parameter Lie point symmetry group. Substituting $\lambda_3=0$ and the actual forms of $G_2$ and $G_3$ in Eq. (\ref{d12}) and integrating, we obtain
\begin{equation}
g(x)=g_{1}e^{-\int{fdx}}{\Bigl(\lambda_1+\int{e^{\int{fdx}}}dx\Bigl)}^{1-\lambda_2},\,\,\lambda_2\neq4,\label{d21}
\end{equation}
so that Eq. (\ref{intr2}) takes the form
\begin{eqnarray}
 \ddot{x}+f(x)\dot{x}^2+g_{1}e^{-\int{fdx}}{\Bigl(\lambda_1+\int{e^{\int{fdx}}}dx\Bigl)}^{1-\lambda_2}=0,\,\,\,\lambda_2\neq4.\label{d22}
\end{eqnarray}
Note that in Eq. (\ref{d21}), for the choice $\lambda_2=1$ we have $g(x)=g_1e^{-\int{fdx}}$, where $g_1$ is an arbitrary constant. This is exactly the same case (\ref{c10}) with $g_1=0$ ($g_2$ is an arbitrary constant which can be considered as $g_1$ for analogy) admitting eight Lie point symmetries. So we exclude the $\lambda_2=1$ case in Eq.(\ref{d22}) from the present two parameter symmetry case. 

The infinitesimal symmetries associated with Eq. (\ref{d22}) are 
\begin{eqnarray}
\xi=a_1+a_2t,\,\,
\eta=\frac{2a_2}{\lambda_2}e^{-\int{fdx}}\Bigl(\lambda_1+\int{e^{\int{fdx}}}dx\Bigl).\label{d23}
\end{eqnarray}
The corresponding infinitesimal generators become
\begin{eqnarray}
X_1=\frac{\partial}{\partial t},\,X_2=t\,\frac{\partial}{\partial t}+\frac{2}{\lambda_2}e^{-\int{fdx}}\Bigl(\lambda_1+\int{e^{\int{fdx}}}dx\Bigl)\,\frac{\partial}{\partial x}. \label{d24}
\end{eqnarray}
satisfying the commutation relation
\begin{eqnarray}
[X_1,X_2]=X_1.\label{d25}
\end{eqnarray}

\subsubsection*{Example:}
As an example for the two parameter symmetry group case, we consider the nonlinear ODE (for $\lambda_2=3$)
\begin{equation}
 \ddot{x}-\frac{2}{x}\,\dot{x}^2+\frac{\lambda x^4}{(\lambda_1x-1)^2}=0.\label{dEx6}
\end{equation}
The infinitesimal symmetries of Eq. (\ref{dEx6}) can be written from Eq. (\ref{d23}) as
\begin{equation}
 \xi=a_1+a_2t,\,\,
\eta=\frac{2a_2}{3}(\lambda_1\,x^2-x).\label{dEx7}
\end{equation}
Then infinitesimal generators can be written as
\begin{eqnarray}
X_1=\frac{\partial}{\partial t},\,
X_2=t\,\frac{\partial}{\partial t}+\frac{2}{3}\,(\lambda_1\,x^2-x)\,\frac{\partial}{\partial x}.\label{dEx8}
\end{eqnarray}
The solution of (\ref{dEx6}) can be written in an implicit form
\begin{eqnarray}
\lambda \,\lambda_1^{3/2}\,x\ln\Bigl(\frac{2\,\sqrt{m}}{\lambda \,\lambda_1^{3/2}\,x}\Bigl(m-\lambda_1\,x(m+\lambda)-\sqrt{m(\lambda_1x-1)(I_1\lambda_1^2x-1)}\,\Bigl)\Bigl)\nonumber\\
-\sqrt{\lambda_1m(\lambda_1x-1)(I_1\lambda_1^2x-m)}\pm m^{3/2}(t-t_0)x=0,
\end{eqnarray}
where $I_1$ and $t_0$ are integration constants and $m=I_1\lambda_1-2\lambda$.

\subsection{One parameter symmetry ($c\,G_2+d\,G_3=0$)}
Next we consider the case $c\,G_2+d\,G_3=0$. As $G_2\neq0$ and $G_3\neq0$ we have necessarily $c=0$ and $d=0$. With this set of restrictions in (\ref{a17.21}) and (\ref{a17.22}) we get $\dot{a}=0,\,g\neq0$, which gives $a=a_1$, where $a_1$ is an arbitrary constant. Obviously this leads to an one parameter symmetry group. It corresponds to time translation generator $X=\frac{\partial}{\partial t}$ irrespective of the form of $f$ and $g$. Hence all other forms of $f$ and $g$ which do not belong to eight, three, and two parameter symmetry groups belong to the family of one parameter symmetry group. 

Now, multiplying with an integrating factor $\dot{x}e^{2\int{f(x)dx}}$, Eq. (\ref{intr2}) can be written as 
\begin{equation}
\dot{x}\,e^{2\int f(x)dx}\left(\ddot{x}+f(x)\,\dot{x}^2\right)=-g(x)\,\dot{x}\,e^{2\int f(x)dx}\label{m1}
\end{equation} 
After an integration the above equation can be brought to the form
\begin{equation}
\dot{x}^2e^{2\int f(x)dx}+2\int g(x)\,e^{2\int f(x)dx}dx=I_1,\label{m2}
\end{equation}
where $I_1$ is an integration constant. A second integration leads to the quadrature
 \begin{eqnarray}
 \pm (t-t_0)=\int {\frac{dx}{\sqrt{e^{-2\int {f(x)\,dx}}(I_1-2\int {g(x)e^{2\int {f(x)\,dx}}}dx)}}},\label{m3}
\end{eqnarray}
where $t_0$ is the second integration constant. Depending on the form of $f(x)$ and $g(x)$, one may or may not be able to carry out the integration on the right hand side of (\ref{m3}) explicitly. We now consider two specific examples belonging to this class.

\subsubsection*{Example 1: Mathews-Lakshmanan Oscillator \cite{mathews}}
The equation of motion is 
\begin{equation}
\ddot{x}-\frac{\lambda \,x}{1+\lambda \,x^2}\dot{x}^2+\frac{\omega_0^2\,x}{1+\lambda \,x^2}=0,\label{m4}
\end{equation}
with $f(x)=-\frac{\lambda \,x}{1+\lambda \,x^2}$ and $g(x)=\frac{\omega_0^2\, x}{1+\lambda \,x^2}$. It admits only the translational symmetry as point symmetry. Using these forms of $f$ and $g$ in Eq. (\ref{m3}), we can write down the general solution as \cite{mathews}
\begin{equation}
x(t)=A\sin(\Omega t+\delta),\,\,\,\Omega=\frac{\omega_0}{\sqrt{1+\lambda A^2}}.\label{m5}
\end{equation}
Eq. (\ref{m4}) has a Hamiltonian
\begin{equation}
H=\frac{1}{2}\,p^2(1+\lambda \,x^2)+\frac{1}{2}\,\frac{\omega_0^2\,x^2}{(1+\lambda \,x^2)},\label{m6}
\end{equation}
where the canonical conjugate momentum $p=\frac{\dot{x}}{1+\lambda x^2}$. Note that when $\lambda $ is negative, $\mid x \mid<\frac{1}{\sqrt{\lambda}}$. The corresponding phase space structure for Eq. (\ref{m4}) is given in Fig. 3.
\begin{figure}[!ht]
\begin{center}
\includegraphics[width=.3\linewidth]{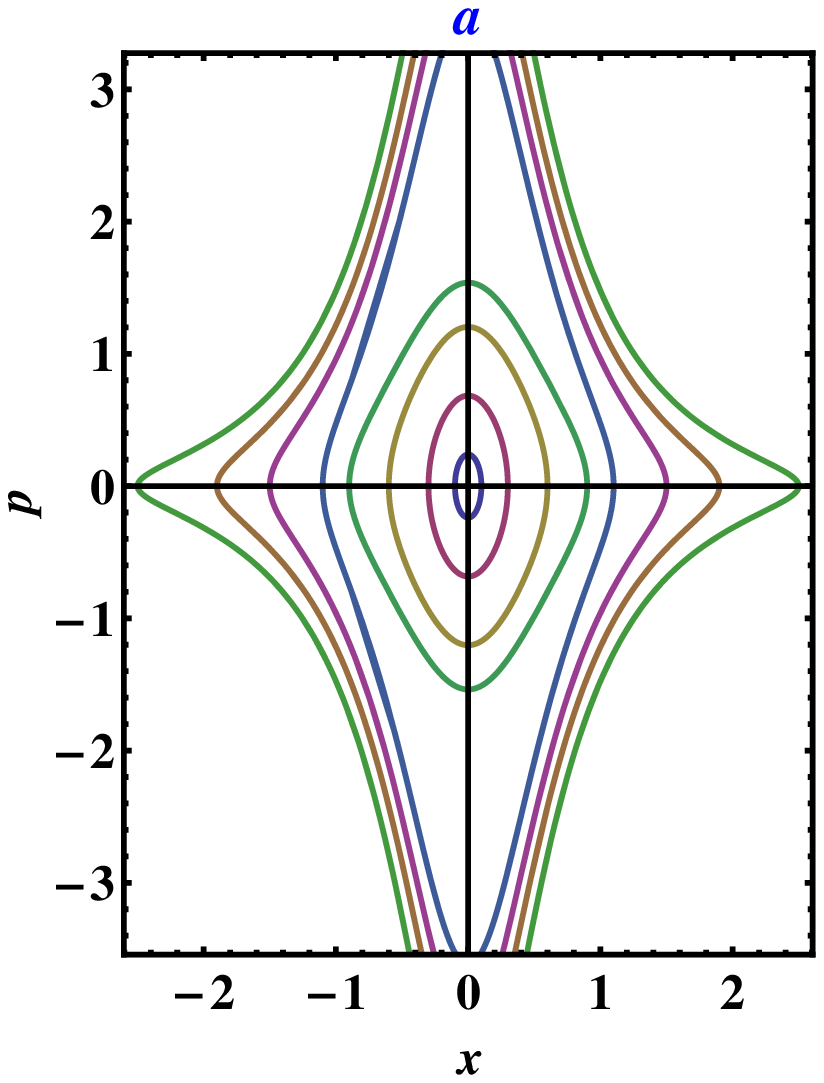}
\includegraphics[width=.265\linewidth]{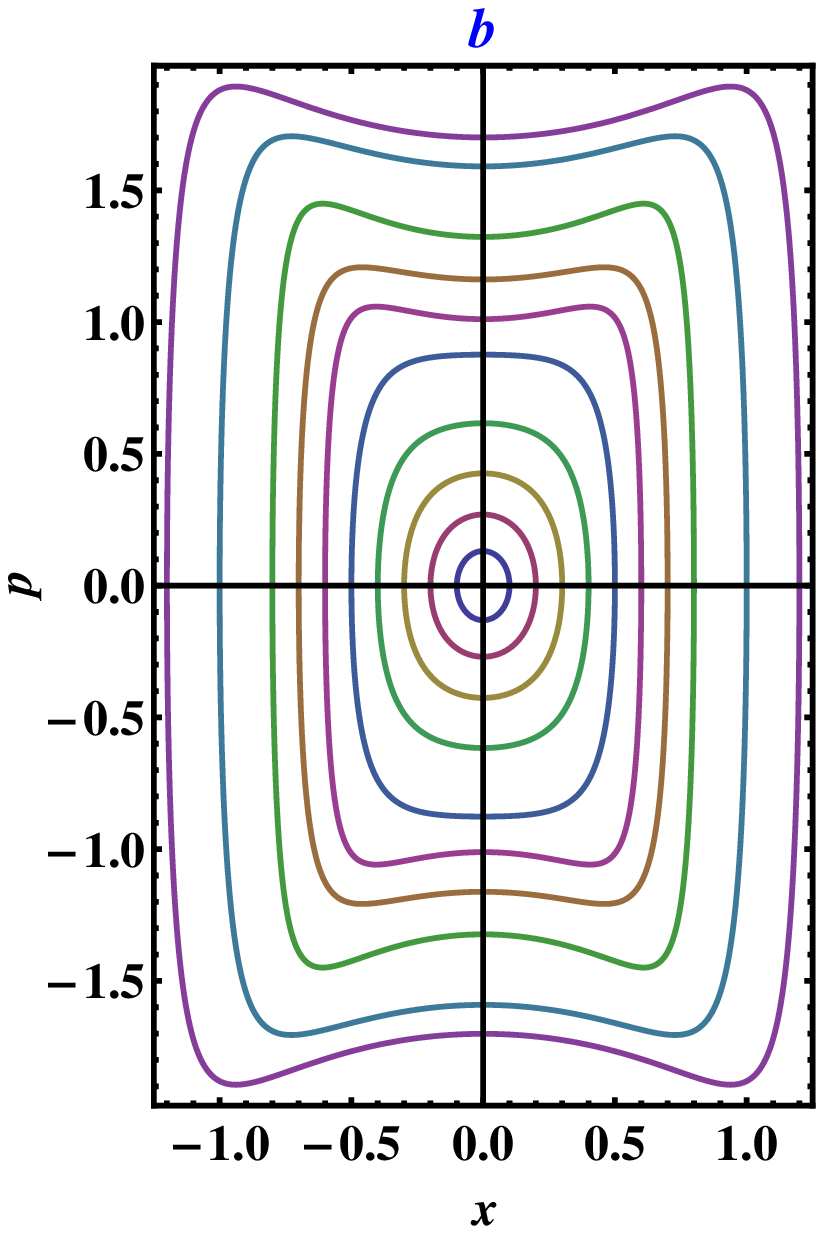}
\caption{(Color online) Phase portrait of equation (\ref{m4}) for different values of ${A}$ corresponding to the Hamiltonian (\ref{m6}), (a)\,$\lambda>0$ and\,(b)\,$\lambda<0$ with the restriction $\mid\, x \,\mid<\frac{1}{\sqrt{\lambda}}$.}
\end{center}
\end{figure}

\subsubsection*{Example 2: Particle in a rotating parabolic well \cite{ml1}}
Similarly a particle moving in a rotating parabolic potential well having the equation of motion
\begin{equation}
\ddot{x}+\frac{\lambda \,x}{1+\lambda \,x^2}\,\dot{x}^2+\frac{\omega_0^2\,x}{1+\lambda \,x^2}=0,\,\, \lambda>0\label{m7}
\end{equation}
is a well known model \cite{mathews,ml1,nay}. The underlying Hamiltonian is found to be 
\begin{equation}
H=\frac{1}{2}\,\frac{p^2}{(1+\lambda \,x^2)}+\frac{1}{2}\,\omega_0^2\,x^2,\label{k6}
\end{equation}
and the canonical momentum
\begin{eqnarray}
p=\dot{x}\,(1+\lambda \,x^2).
\end{eqnarray}
Eq. (\ref{m7}) admits only the translational symmetry as the admissible Lie point symmetry. Using Eq. (\ref{m2}) one can write the first integral for Eq. (\ref{m7}) as
\begin{eqnarray}
(1+\lambda \,x^2)\,\dot{x}^2+\omega_0^2\,x^2=I_1.\label{z1}
\end{eqnarray}
The above equation can be integrated to get the solution in terms of complicated elliptic integral of second kind.

Considering the general equation corresponding to one parameter Lie point symmetry, as noted above, one can deduce the first (energy) integral from the time translational symmetry which in turn can be integrated to yield the general solution. One can also rewrite the first integral of the above such examples as a linear first-order equation in some new variables, which can be integrated to give the second integral explicitly. Now, the question arises why the above mentioned examples are integrable explicitly when they admit only one Lie point symmetry. Recently, Bhuvaneswari et al. \cite{bhu} proved that Eq. (\ref{m4}) exhibits the so called nonlocal $\lambda$-symmetries as the possible reason for the integrability of such systems. Hence one can explain the integrability of such systems by associating more general symmetries to them. So there exists the question of identifying the systems belonging to the family of one parameter Lie point symmetry that are explicitly integrable. This problem has not been pursued in this study and will be discussed separately. 

\section{Lie symmetries of Eq. (\ref{intr2}) with $f(x)=0$ or $g(x)=0$}
\label{sec7}
If we consider the special case of Eq. (\ref{intr2}) with $f(x)=0$, then we have
\begin{equation}
 \ddot{x}+g(x)=0.\label{e1}
\end{equation}
The above equation has been studied in detail by Pandey \textit{et al.} \cite{snp1} and it has been shown that it admits two and three parameter symmetry generators for certain forms of $g(x)$, while for other choices it admits only one parameter symmetry group. 

Next, for the case $g(x)=0$, Eq. (\ref{intr2}) takes the form
\begin{equation}
 \ddot{x}+f(x)\dot{x}^2=0.\label{e2}
\end{equation}
To get the symmetry generators of Eq. (\ref{e2}) we substitute $g$ as zero in Eqs. (\ref{a16.3}) and (\ref{a16.4}) and $G_{1x},\,G_{2x}$ 
and $G_{3x}$ in terms of $G_1,\,G_2$ and $G_3$ and then equate the coefficients of $G_1,\,G_2$ and $G_3$ to zero. Doing so we arrive at the following determining equations for the infinitesimals, that is
\begin{eqnarray}
\ddot{b}=0,\quad \ddot{c}=0,\quad \ddot{d}=0,\quad \ddot{a}-2\dot{c}=0.\label{e3}
\end{eqnarray}
Solving these equations one can get the forms of $a,b,c$ and $d$ which on substituting in Eq. (\ref{a12}) and (\ref{a16}) give the symmetry generators as
\begin{eqnarray}
 \xi&=&a_1+a_2t+c_1t^2+(b_1t+b_2)\,\Im, \nonumber\\
\eta&=&2\,b_1 G_1+(c_1t+c_2) G_2+(d_1t+d_2)G_3,
\end{eqnarray}
where $\Im,G_1,G_2$ and $G_3$ are already defined in Sec. II. Here $a_1,a_2,b_1,b_2,c_1,c_2,d_1$ and $d_2$ are the eight symmetry generators. \\
The corresponding infinitesimal generators can be then written as
\begin{eqnarray}
X_1&=&\frac{\partial}{\partial t},\,\,\,X_2=t\,\frac{\partial}{\partial t},\,\,\,X_3=t\,\Im\,\frac{\partial}{\partial t}+2\,G_1\frac{\partial}{\partial x},\,\,\,X_4=\,\Im\,\frac{\partial}{\partial t},\nonumber\\
X_5&=&t^2\,\frac{\partial}{\partial t}+t\,G_2\,\frac{\partial}{\partial x},\,\,\,X_6=G_2\,\frac{\partial}{\partial x},\,\,\,X_7=t\,G_3\,\frac{\partial}{\partial x},\,\,\,X_8=G_3\,\frac{\partial}{\partial x},
\end{eqnarray}
satisfying the $sl(3,R)$ algebra. Obviously Eq. (\ref{e2}) can be linearized under the transformation $X=h(x)$ to the free particle 
oscillator equation
\begin{equation}
\ddot{X}=0,
\end{equation}
where $h(x)=h_1\int{e^{\int{f(x)dx}}dx}+h_2$, $h_1$ and $h_2$ being arbitrary constants.

\section{Equivalence transformations}
\label{sec8}
Finally we consider equivalence transformations \cite{lv} associated with (\ref{intr2}). Let us consider a set of smooth, locally one-to-one transformations $\mathcal{T}: (t,x,f,g) \longrightarrow(T,X,f_1,g_1) $ of the space $\mathbb{R}^4$ that act by the formulae
\begin{align}
T=G(t,x),\;\; X=F(t,x), \;\; f_1=H(t,x,f),\;\; g_1=L(t,x,g).
\label{f1}
\end{align}

An equivalent transformation of Eq. (\ref{intr2}) is an invertible transformation that converts Eq. (\ref{intr2}) to an equation of the same form \cite{lv}
\begin{align}
\ddot{X}=-f_1(X)\dot{X}^2-g_1(X).
\label{f3}
\end{align}
In this case Eqs. (\ref{intr2}) and (\ref{f3}) and the functions $\{f(x)$,$g(x)\}$ and
$\{f_1(X)$,$g_1(X)\}$ are equivalent.

Substituting the transformation (\ref{f1}) into Eq. (\ref{f3}) we get
\begin{align}
& f_1(F_t+\dot{x}F_x)^2(G_t+\dot{x}G_x)+g_1(G_t+\dot{x}G_x)^3= (F_t+\dot{x}F_x)[(G_{tt}+2\dot{x}G_{tx}
\nonumber\\
 & \qquad
+\dot{x}^2G_{xx}-G_x(f\dot{x}^2+g)]-(G_t+\dot{x}G_x)[(F_{tt}+2\dot{x}F_{tx}+\dot{x}^2F_{xx}-F_x(f\dot{x}^2+g)],
\label{f4}
\end{align}
where the subscripts denote partial derivative with respect to that variable. Equating the coefficients of different powers of $\dot{x}^n,\;n=0,1,2,3$, we get
\begin{align}
f_1F_x^2G_x+g_1G_x^3 & =  F_xG_{xx}-G_xF_{xx},\label{f5a}\\
f_1(2F_xF_tG_x+F_x^2G_t)+3g_1G_x^2G_t  & =  Jf+F_tG_{xx}- G_tF_{xx}+2F_xG_{tx}-2G_xF_{tx},\label{f5b}\\
 f_1(2F_xF_tG_t+F_t^2G_x)+3g_1G_t^2G_x & = F_xG_{tt}-G_xF_{tt}+2F_tG_{tx}-2G_tF_{tx},\label{f5c}\\
 f_1F_t^2G_t+g_1G_t^3 & =  G_{tt}F_{t}-G_tF_{tt}+gJ,\label{f5d}
\end{align}
where $J=G_tF_x-F_tG_x$. Solving Eqs. (\ref{f5a}) and (\ref{f5b}) we get the forms of the functions $f_1$ and $g_1$ as
\begin{eqnarray}
&&f_1=-\frac{2G_tG_xF_{xx}-3F_xG_tG_{xx}+fG_xJ+F_tG_xG_{xx}+2F_xG_xG_{tx}-2G_x^2F_{tx}}{2F_xG_xJ}\label{f6},\\
&&g_1=-\frac{F_xG_{xx}(G_tF_x+F_tG_x)+F_xG_x(2(G_xF_{tx}-F_xG_{tx})-fJ)-2G_x^2F_tF_{xx}}{2G_x^3J}\label{f7}.
\end{eqnarray}
Substituting these values of $f_1$ and $g_1$ in Eqs. (\ref{f5c}) and (\ref{f5d}) and then equating both the equations thus obtained, we get a general condition on the forms of the functions $F$ and $G$ as
\begin{eqnarray}
&&(G_t-1)[(J+2f_tG_x)(2G_x^2F_{tx}+JG_{xx})-G_x(2G_x(F_tG_tF_{xx}+F_xG_xF_{tt})+F_xG_t(2F_xG_{tx}\nonumber\\
&&+fJ))]+(F_tG_x-F_x)[2G_t^2F_xG_{xx}+2G_x^2F_xG_{tt}-2F_xG_xG_tG_{tx}-fJG_xG_t]+J[2F_xG_xG_{tx}\nonumber\\
&&+f(JG_x+2G_x^2F_t)-2gF_xG_x^3]=0.\label{f8}
\end{eqnarray}
The above equation is difficult to solve for the general forms of $F$ and $G$. Hence one can consider specific forms of $F$ and $G$ for simplicity. One of the possible solutions for the set of Eqs. (\ref{f5a})-(\ref{f5d}) is of the form
\begin{eqnarray}
F=\alpha x+\beta, \,\,\,G=\gamma t+\delta, \label{f9}
\end{eqnarray}
where $\alpha,\beta,\gamma$ and $\delta$ are arbitrary constants. Substituting the above forms of $F$ and $G$ in Eqs. (\ref{f5a})-(\ref{f5d}), we get the forms for the functions $f_1$ and $g_1$ as
\begin{eqnarray}
f_1=\frac{f}{\alpha},\,\,g_{1}=\frac{\alpha g}{\gamma^2}.
\end{eqnarray}
Then the possible equivalence transformation is
\begin{eqnarray}
&&X=\alpha x+\beta, \,\,\,G=\gamma t+\delta,\\
&&f_1=\frac{f}{\alpha},\,\,g_{1}=\frac{\alpha g}{\gamma^2}.
\end{eqnarray}
Making use of the above, we can write down the equivalence transformation for some of the examples discussed earlier. For example, for Eq. (\ref{dEx1.1}) the transformation is given as 
\begin{eqnarray}
X&=&\alpha x+\beta, \,\,\,G=\gamma t+\delta,\\
f_1&=&-\frac{2}{X-\beta},\,\,g_1=\frac{\lambda (X-\beta)^5}{\alpha \gamma^2(\lambda_1(X-\beta)-\alpha)^3}\,,
\end{eqnarray}
whereas for Eq. (\ref{dEx6}) the transformation is
\begin{eqnarray}
X&=&\alpha x+\beta, \,\,\,G=\gamma t+\delta,\\
f_1&=&-\frac{2}{X-\beta},\,\,g_1=\frac{\lambda (X-\beta)^4}{\alpha \gamma^2(\lambda_1(X-\beta)-\alpha)^2}\,.
\end{eqnarray}
One can analyse Eq. (\ref{f8}) further to get more general equivalence transformations. We have not pursued this problem further in the present work.

\section{Conclusion}
\label{sec9}
In this paper we have investigated systematically the Lie point symmetry groups associated with the quadratic Li$\acute{e}$nard Eq. (\ref{intr2}). Even though the integrability properties of the general form of Eq. (\ref{intr2}) has been discussed to some extent in the literature, we have systematically identified and classified all those equations which admit one, two, three and eight parameter symmetry groups. We have found the general form of (\ref{intr2}) that belongs to the linearizable case as the one given by Eq. (\ref{b3}) admitting the maximal (eight parameter) symmetry group, whereas for the integrable but not linearizable cases the general forms for three and two parameter symmetry groups are represented by Eqs. (\ref{d16}) and (\ref{d22}), respectively. We have also deduced the interesting result that the condition for isochronicity of Eq. (\ref{intr2}) is the same as that of the linearizability condition. Our analysis clearly confirm the powerful nature of analysis of nonlinear ODEs based on symmetry properties.

\section{Acknowledgments}
AKT and SNP are grateful to the Centre for Nonlinear Dynamics, Bharathidasan University, Tiruchirappalli, for warm hospitality. The work of SNP forms part of a Department of Science and Technology, Government of India, sponsored research project. The work of MS forms part of a research project sponsored by UGC. The work forms part of a Department of Science and Technology, Govt. of India IRHPA project and a Ramanna Fellowship project of ML. He also acknowledges the financial support provided through a DAE Raja Ramanna Fellowship.

\end{document}